\documentclass[twocolumn,showpacs,preprintnumbers,prd,superscriptaddress,nofootinbib]{revtex4-1}

\usepackage[colorlinks=true,citecolor=blue,urlcolor=blue]{hyperref}
\usepackage{amsfonts}
\usepackage{amsmath}
\usepackage{amssymb}
\usepackage{graphicx}
\usepackage{bm}
\usepackage{color}
\usepackage{comment}
\usepackage{tikz}

\definecolor{lime}{HTML}{A6CE39}
\DeclareRobustCommand{\orcidicon}{%
	\begin{tikzpicture}
	\draw[lime, fill=lime] (0,0) 
	circle [radius=0.16] 
	node[white] {{\fontfamily{qag}\selectfont \tiny ID}};
	\draw[white, fill=white] (-0.0625,0.095) 
	circle [radius=0.007];
	\end{tikzpicture}
	\hspace{-2mm}
}

\foreach \x in {A, ..., Z}{%
	\expandafter\xdef\csname orcid\x\endcsname{\noexpand\href{https://orcid.org/\csname orcidauthor\x\endcsname}{\noexpand\orcidicon}}
}

\begin{document}

\title{Testing  Yukawa cosmology at the Milky Way and M31 galactic scales}

\author{Rocco  D'Agostino\orcidA{}}
\email{rocco.dagostino@unina.it}
\affiliation{Scuola Superiore Meridionale,  Largo San Marcellino 10, 80138 Napoli, Italy.}
\affiliation{Istituto Nazionale di Fisica Nucleare, Sezione di Napoli, Via Cinthia, 80126 Napoli, Italy.}

\author{Kimet Jusufi\orcidB{}}
\email{kimet.jusufi@unite.edu.mk}
\affiliation{Physics Department, State University of Tetovo, Ilinden Street nn, 1200, Tetovo, North Macedonia.}

\author{Salvatore Capozziello\orcidC{}}
\email{capozziello@unina.it}
\affiliation{Dipartimento di Fisica ``E. Pancini", Universit\`a di Napoli ``Federico II", Via Cintia, 80126 Napoli, Italy.}
\affiliation{Scuola Superiore Meridionale,  Largo San Marcellino 10, 80138 Napoli, Italy.}
\affiliation{Istituto Nazionale di Fisica Nucleare, Sezione di Napoli, Via Cinthia, 80126 Napoli, Italy.}

\begin{abstract}
We address the galaxy rotation curves through the Yukawa gravitational potential emerging as a correction of the Newtonian potential in extended theories of gravity. On the one hand, we consider the contribution of the galactic bulge, galactic disk, and the dark matter halo of the Navarro-Frenk-White profile, in the framework of the standard $\Lambda$CDM model. On the other hand, we use modified Yukawa gravity to show that the rotational velocity of galaxies can be addressed successfully without the need for dark matter. In Yukawa gravity, we recover MOND and show that dark matter might be seen as an apparent effect due to the modification of the law of gravitation in terms of two parameters: the coupling constant $\alpha $ and the characteristic length $\lambda$. 
We thus test our theoretical scenario using the Milky Way and M31 rotation velocity curves. In particular, we place observational constraints on the free parameters of Yukawa cosmology through the Monte Carlo method and then compare our results with the predictions of the $\Lambda$CDM paradigm by making use of Bayesian information criteria.
Specifically, we find that $\lambda$ is constrained to be of the order of kpc, while cosmological data suggest $\lambda$ of the order of Gpc. To explain this discrepancy, we argue that there is a fundamental limitation in measuring $\lambda$ due to the role of quantum mechanics on cosmological scales. 
\end{abstract}

\maketitle

\section{Introduction}
According to present observations, the best picture of cosmology suggests that our universe is uniform and isotropic at large scales. However, one of the most interesting discoveries is the presence of cold dark matter, a mysterious form of matter that interacts only via the gravitational force \cite{Peebles:1984zz,Bond:1984fp,Trimble:1987ee,Turner:1991id}. Despite numerous efforts, there has been no direct detection of dark matter particles, and their existence is solely manifested through the gravitational impacts they exert on galaxies and larger cosmic structures \cite{Salucci:2020nlp}. Furthermore, dark energy has been introduced as an explanatory concept to account for the universe's accelerating expansion pointed out by numerous observations, and it is linked to the cosmological constant \citep{Carroll:1991mt,SupernovaCosmologyProject:1997zqe, SupernovaSearchTeam:1998fmf,Peebles:2002gy}.

The theoretical scenario describing the aforementioned cosmological features is the $\Lambda$CDM paradigm, which stands as the most successful model in modern cosmology. This framework adeptly accounts for a wide range of cosmological observations while employing a minimal set of parameters \cite{Planck:2018vyg}. Nevertheless, fundamental issues remain connected to the deep understanding of the nature and behavior of both dark matter and dark energy \cite{Weinberg:1988cp,Zlatev:1998tr,Padmanabhan:2002ji}. This knowledge gap persists also when scalar fields are invoked to play a key role in the physical depiction of the universe, as in the context of the inflationary scenario \citep{Starobinsky:1980te,Guth:1980zm,DAgostino:2021vvv,DAgostino:2022fcx, Capozziello:2022tvv}. 
On the other hand, recent inconsistencies among cosmological datasets have highlighted some tensions inherent to the standard $\Lambda$CDM picture, questioning the accuracy of the model itself to describe the entire universe evolution and dynamics \cite{DiValentino:2020zio,Riess:2021jrx,Perivolaropoulos:2021jda,DAgostino:2023cgx}.

All the above problems motivated the studies of  different perspectives  advocating for the possibility of explaining observational data through modifications of Einstein equations, resulting in alternative or extended theories of gravity \cite{Capozziello:2002rd,CANTATA:2021ktz,DeFelice:2010aj,Nojiri:2010wj,Clifton:2011jh,Capozziello:2011et,Bahamonde:2015zma,BeltranJimenez:2017tkd,Capozziello:2019cav,DAgostino:2019hvh,Koyama:2015vza,DAgostino:2022tdk,Capozziello:2022rac,Joyce:2016vqv,Bajardi:2022tzn,Capozziello:2023ccw}. In the context of dark matter, which is usually assumed to explain the flatness of galaxy rotation curves \cite{Salucci:2018eie}, one of the initial theories put forward to address this phenomenon was the Modified Newtonian Dynamics (MOND), first introduced in Ref.~\cite{Milgrom:1983ca}, involving modifications of the Newton law
\cite{Ferreira:2009eg, Milgrom:2003ui, Tiret:2007kq, Kroupa:2010hf, Cardone:2010ru, Richtler:2011zk, Bekenstein:2004ne}.
Other intriguing proposals in this domain include superfluid dark matter \cite{Berezhiani:2015bqa} and the concept of Bose-Einstein condensate \cite{Boehmer:2007um}, among others. These ideas represent diverse approaches to addressing the challenges posed by dark matter's role in galactic rotation curves.

The present work is devoted to studying the Yukawa potential in galactic systems in view to elucidate the nature of dark matter. In this model, dark matter is postulated to be explicable through the coupling between baryonic matter mediated by a long-range force, represented by the Yukawa gravitational potential. The Yukawa model is described by two crucial parameters: the coupling parameter $\alpha$ and the effective length parameter $\lambda$, which is related to the graviton mass. The Yukawa potential can appear in numerous scenarios, including $f(R)$ gravity \cite{Capozziello:2007ms,Capozziello:2014mea,DeMartino:2017ztt,Benisty:2023ofi}. As  recently shown  in Refs.~\cite{Jusufi:2023xoa,Gonzalez:2023rsd}, using the Yukawa potential one can obtain the $\Lambda$CDM as an effective model. In this picture, the dark matter is only an apparent effect that naturally appears from the long-range force associated with the graviton. This modifies Einstein's gravity at large distances, and the amount of dark matter follows from the distribution of baryonic matter undergoing the Yukawa-like gravitational interaction \cite{Capozziello:2011et,Jusufi:2023xoa,Gonzalez:2023rsd}. 
In this perspective, one of the most famous evidence for dark matter is the rotation curves of galaxies. In the present paper, we aim to test and study in more detail the Yukawa gravitational potential for rotating curves. In particular, we aim to show how the Yukawa potential can explain the rotating curves without the need for dark matter. The latter, in fact, appears as an effect of the modification of the law of gravitation. 

The paper is organized as follows. In Section~\ref{sec:f(R)}, we review the Yukawa potential obtained from $f(R)$ gravity. 
In Section~\ref{sec:cosmology}, we examine the modified Friedmann equations in Yukawa cosmology while, in Section~\ref{sec:rotation}, we address the problem of rotating galaxy curves in the framework of the $\Lambda$CDM model and the new cosmological scenario. 
In Section~\ref{sec:constraints}, we use astrophysical data of the Milky Way (MW) and M31 galaxies to constrain the free parameters of the  Yukawa model. 
Furthermore, in Section~\ref{sec:discussion}, we discuss  results in light of the most recent findings in the literature. 
Finally, Section~\ref{sec:conclusion} is devoted to the conclusions and final remarks.

Throughout this work, we use natural units of $c=\hbar=1$, unless otherwise specified.

\section{Yukawa potential in $f(R)$ gravity}
\label{sec:f(R)}

Yukawa-like corrections to the Newtonian potential naturally emerge in the weak field limit of extended theories of gravity, such as  $f(R)$ models \cite{Capozziello:2002rd,Sotiriou:2008rp,DeFelice:2010aj}. In particular, the $f(R)$ gravity action can be written as 
\begin{equation}
S=\dfrac{1}{16\pi G}\int d^{4}x\,\sqrt{-g}\,f(R)+S_{\rm matter}[g_{\mu \nu},\Phi_i]\,,
\label{eq:f(R) FE}
\end{equation}
where $R$ is the Ricci scalar, $G$ is the Newton gravitational constant and $g$ is the determinant of the metric tensor, $g_{\mu\nu}$. In addition $\Phi_i$ are matter fields.
By varying the above action with respect to $g_{\mu\nu}$, we obtain the field equations
\begin{equation}
f'(R)R_{\mu\nu}-\frac{1}{2}f(R)g_{\mu\nu}-f'(R)_{;\mu\nu}+g_{\mu\nu}\Box f'(R)= 8\pi G\, T_{\mu\nu}\,,
\end{equation}
where $T_{\mu\nu}$ is the matter energy-momentum tensor. 
Here, the semicolon and the prime denote the covariant derivative and the derivative with respect to $R$, respectively, while $\Box$ stands for the d'Alembert operator. 
Taking the trace of Eq.~\eqref{eq:f(R) FE}, one finds
\begin{equation}\label{fetr}
3\Box f'(R)+f'(R)R-2f(R)= 8\pi G\, T\,,
\end{equation}
It is worth noticing that Einstein's equations of general relativity are recovered in the limit $f(R)\rightarrow R$.

To show how the Yukawa-like potential emerges in $f(R)$ theories, we take into account the corresponding field equations in the presence of matter. In the weak field limit, we can perturb the metric tensor as 
\begin{equation}
g_{\mu\nu}\,=\,\eta_{\mu\nu}+h_{\mu\nu}\,,   
\end{equation}
where $|h_{\mu\nu}|\ll \eta_{\mu\nu}$ is a small perturbation around the Minkowsky spacetime, $\eta_{\mu\nu}$. 
Assuming $f(R)$ to be analytic, one can consider the Taylor series \cite{Capozziello:2007ms,Cardone:2011ze,Capozziello:2020dvd,Benisty:2023ofi}
\begin{equation}
f(R)\simeq f(R_0)+f'(R_0)(R-R_0)+f''(R_0)(R-R_0)^2/2\,.  
\end{equation}
Then, imposing a spherical symmetry, we have
\begin{equation}
\label{Edd}
    g_{00}=-\left(1-2\Phi(r)\right),
\end{equation}
leading to the gravitational Yukawa-like potential
\begin{eqnarray}
\Phi(r) = -\frac{G M}{r}\left(\frac{1+\alpha \,   e^{-r/\lambda}}{1 + \alpha}\right),
\end{eqnarray}
with $\alpha = f'(R_0) - 1$ 
and $\lambda^2=-(1+\alpha)/(6f''(R_0))$ can be interpreted as the scale length of the interaction due to the graviton.
Under the rescaling $G \to G (1+\alpha)$, we finally have
\begin{eqnarray}
\Phi(r) = -\frac{G M}{r} \left(1+\alpha \,   e^{-m r}\right),
\end{eqnarray}
where $m=1/\lambda$ represents the graviton mass. We notice that the Newtonian potential is recovered for $\alpha=0$, that is for $f(R)\rightarrow R$\footnote{We refer the reader to Refs.~\cite{Capozziello:2007id, Capozziello:2009vr,Napolitano:2012fp} for a detailed discussion on the value and the sign of the parameter $\alpha$ to be consistent with observations.}.

\section{Yukawa cosmology}
\label{sec:cosmology}

The Yukawa potential may be conveniently expressed in terms of an effective length which can be considered as the wavelength of a massive graviton. The Yukawa gravitational potential was modified via the quantum deformed parameter $l_0$, in the following form \cite{Jusufi:2023xoa}:
\begin{equation}
\Phi(r)=-\frac{G M m}{\sqrt{r^2+l_0^2}}\left(1+\alpha\,e^{-\frac{r}{\lambda}}\right).
\end{equation}
However, $l_0$ is important only in the early universe \cite{Jusufi:2022mir}, meaning that we can set $l_0/r \to 0$ in the late-time universe. Using the relationship $F=-\nabla \Phi(r)$, we find the correction to Newton's law of gravitation: 
\begin{equation}\label{F5}
F=-\frac{G M m}{r^2}\left[1+\alpha\,\left(\frac{r+\lambda}{\lambda}\right)e^{-\frac{r}{\lambda}}\right].
\end{equation}
Let us assume now the spacetime background as characterized by the Friedmann-Robertson-Walker (FRW) metric:
\begin{equation}
ds^2=-dt^2+a^2(t)\left[\frac{dr^2}{1-kr^2}+r^2(d\theta^2+\sin^2\theta
d\phi^2)\right],
\end{equation}
where $\mathcal{R}=a(t)r$ is the apparent FRW horizon radius, with $a(t)$ being the normalized scale factor as a function of cosmic time, $t$. Here, the spatial curvature parameter $k=\{0,1,-1\}$ describes a flat, closed and open universe, respectively. 
If we consider a matter source to be modeled as a perfect fluid with energy density $\rho$ and pressure $p$, one can write the energy-momentum tensor as
\begin{equation}\label{T}
T_{\mu\nu}=(\rho+p)u_{\mu}u_{\nu}+pg_{\mu\nu}\,,
\end{equation}
along with the continuity equation 
\begin{equation}
\dot{\rho}+3H(\rho+p)=0\,,
\end{equation}
with $H\equiv\dot{a}/a$ being the Hubble parameter. 
Therefore, the first Friedmann equation reads \citep{Jusufi:2023xoa}
\begin{eqnarray}\notag
H^2+\frac{k}{a^2} &=&  \frac{8\pi G_{\rm eff}
}{3}\sum_i \rho_i -\frac{1}{\mathcal R^2}  \sum_i \Gamma_1(\omega_i)\rho_i  \nonumber \\
&+&\frac{4 \pi G_{\rm eff}}{3}\mathcal R^2 \sum_{i}\Gamma_2(\omega_i)\rho_i, 
\label{Fried01}
\end{eqnarray}
where $G_{\rm eff}= G(1+\alpha)$, along with the definitions 
\begin{align}
\Gamma_1 (w_i ) & \equiv  \frac{4 \pi G_{\rm eff} l_0^2 }{ 3 }\left(\frac{1+3
\omega_i}{1+w_i}\right), \\
    \Gamma_2 (w_i )   & \equiv   \frac{\alpha\, (1+3w_i)}{  \lambda^2 (1+\alpha) (1-3w_i)},
\end{align}
with $w_i=p_i/\rho_i$ being the equation of state parameter of the $i$-th cosmic species. 
Here, $\Gamma_1$ plays an important role in the early universe, while $\Gamma_2$ plays an important role in the late-time universe. We notice that in the last equation, there appears a singularity in the case of radiation $(w=1/3)$. This suggests a phase transition in the early Universe, from radiation to a matter-dominated state \cite{Jusufi:2023xoa}. 
In fact, in a radiation-dominated universe, we have $\Gamma_2=0$, and no singularity is present in the $\Gamma_1$ term. In other words, the effect of $\alpha$ becomes important only after the phase transition from a radiation-dominated to a matter-dominated universe. 

We shall then focus on the late-time Universe, where Yukawa modifications to gravity assume a significant role. In this regime, the first Friedmann equation in the late-time universe reads
\begin{equation}
H^2+\frac{k}{a^2}=\frac{8\pi G_{\rm eff}
}{3}\sum_i \rho_i+\frac{4 \pi G_{\rm eff}}{3}\mathcal R^2 \sum_{i}\Gamma_2(\omega_i)\rho_i\,.
\end{equation}
We consider a non-relativistic matter source for the cosmic fluid, in a flat universe scenario\footnote{It is worth mentioning that, although prompted by the cosmic microwave background observations \cite{Planck:2018vyg}, the flat universe scenario is still under debate, see, e.g., Refs.~\cite{DiValentino:2019qzk,Handley:2019tkm,Glanville:2022xes,DAgostino:2023tgm}.}. In this case, $\mathcal{R}^2=H^{-2}$ and, for the Yukawa cosmology, we have \cite{Jusufi:2024ifp}
\begin{align}\label{eq43B}
  E^2(z)&=\frac{(1+\alpha)}{2}\,\left(\Omega_{B,0} 
(1+z)^{3}+\Omega_{\Lambda,0}\right)\notag \\
   &+\frac{(1+\alpha)}{2}\sqrt{\left(\Omega_{B,0} 
(1+z)^{3}+\Omega_{\Lambda,0}\right)^2+\frac{\Omega^2_{DM}(z)}{{(1+z)^3}}},
\end{align}
where $E(z)\equiv H(z)/H_0$ is the reduced Hubble parameter, with $H_0$ being the Hubble constant and  $z\equiv a^{-1}-1$ is the cosmological redshift, while $\Omega_B$ and $\Omega_\Lambda$ are the normalized density parameters related to baryonic matter and dark energy, respectively.
In particular, it has been shown that the dark matter density parameter can be related to the baryonic matter as \cite{Jusufi:2023xoa}
\begin{equation}\label{eqDM}
    \Omega_{DM}= \frac{\sqrt{2 \alpha \Omega_{B,0}}}{\lambda H_0\,(1+\alpha)} 
\,{(1+z)^{3}}\,,
\end{equation}
where the subscript `0' indicates quantities evaluated at present, namely $z=0$.
This implies that dark matter can be understood as a consequence of the modified Newton law, quantified by $\alpha$ and $\Omega_B$. 
Furthermore, one can find an expression that relates between baryonic matter, effective dark matter, and dark energy:
\begin{equation}\label{DMCC}
   \Omega_{DM}(z)= \sqrt{2\, \Omega_{B,0}  \Omega_{\Lambda,0}}{(1+z)^3}\,.
\end{equation}
where we introduced the definition
\begin{equation}
    \Omega_{\Lambda,0}= \frac{1}{\lambda^2 H^2_0}\frac{\alpha}{(1+\alpha)^2}\,.
\end{equation}
Up to the leading-order terms, one finds 
\begin{eqnarray}
   {E^2(z)}&=&(1+\alpha) \left[\bar{\Omega}_B 
(1+z)^{3}+\Omega_{\Lambda}\right],
\end{eqnarray}
The reduced Hubble parameter is thus obtained as \cite{Gonzalez:2023rsd}
\begin{equation}
{E^{2}(z)}=\left(1+\alpha\right)\left[\left(\Omega_{B,0}+ 
\sqrt{2\Omega_{B,0}\Omega_{\Lambda,0}}\right)\left(1+z\right)^{3} +\Omega_{\Lambda,0}\right],
\label{YukawaLCDMLike}
\end{equation}
while the condition $H(z=0)=H_{0}$ leads to
\begin{equation}\label{alphaconstraint}
1+\alpha=\left[{\Omega_{B,0}+\sqrt{2\Omega_{B,0}\Omega_{\Lambda,0}}+\Omega_{\Lambda,0}}\right]^{-1}\,.
\end{equation}
It is important to stress that one may formally obtain the standard $\Lambda$CDM cosmology under the following definitions:
\begin{align}
    \left(1+\alpha\right)\Omega_{B,0} &  \equiv\Omega_{B,0}^{\Lambda\text{CDM}}, 
\label{OB0def}\\
    \left(1+\alpha\right)\sqrt{2\Omega_{B,0} \Omega_{\Lambda,0}} & 
\equiv\Omega_{DM,0}^{\Lambda\text{CDM}}\,, \label{ODMdef}\\
    \left(1+\alpha\right)\Omega_{\Lambda,0}  & 
\equiv\Omega_{\Lambda,0}^{\Lambda\text{CDM}}\,. \label{OLdef}
\end{align}
In fact, Eq.~\eqref{YukawaLCDMLike} can be recast as
\begin{equation}\label{HLCDM}
   {E^{2}(z)}=\left(\Omega_{B,0}^{\Lambda\text{CDM}}+\Omega_{DM,0}^{\Lambda\text{CDM}}\right)\left(1+z\right)^{3}+\Omega_{\Lambda,0}^{\Lambda\text{CDM}},
\end{equation}
which is related to the Hubble parameter of the $\Lambda$CDM model through $H(z)=H_0\, E(z)$, with  $H_{0}=H_{0}^{\Lambda\text{CDM}}$.

\section{Galaxy rotation curves: MOND from Yukawa cosmology}
\label{sec:rotation}

In what follows, we briefly review galaxy rotation in the $\Lambda$CDM model. We then describe how to obtain the galaxy rotation curve in the framework of Yukawa cosmology. Hence, the Navarro-Frenk-White (NFW) profile for dark matter is analyzed in both contexts.

\subsection{Rotation curves in $\Lambda$CDM }

The galaxy rotation curve in the standard $\Lambda$CDM cosmology can be modeled as detailed in Ref.~\cite{Dai:2022had}. Specifically, the total  squared rotational velocity is given as
\begin{equation}
    v^2(r)=v_b^2(r)+v_d^2(r)+v_h^2(r)\,,
\end{equation}
where the three components on the left-hand side refer to the galactic bulge, disk and the dark halo, respectively.

We assume the galactic bulge to be spherically symmetric with a de Vaucouleur profile \cite{deVaucouleurs:1948lna}, with
surface mass density 
\begin{equation}
	\Sigma_b(r) = \Sigma_{b,0}\, e^{-\kappa \left[\left(\frac{r}{a_b}\right)^{1/4}-1\right]}\,,
\end{equation} 
where $\kappa=7.6695$, and $\Sigma_{b,0}$ is the surface mass density at $r=a_b$. 
Note that de Vaucouleurs profile is a good empirical model describing the surface brightness of a galaxy as a function of distance from the center of the galaxy. As such, it has not been obtained from a fundamental law and it does not depend explicitly on the gravitational potential or modified gravity model in question. By construction, this profile is a special case of the S\'ersic profile, which is characterized by the free parameter $b_n=1.9992n-0.32715$, where $n$ is the Sérsic index. For the de Vaucouleurs profile, we here set $n=4$, hence $\kappa=b_4=7.66925$.
The mass density of the bulge is then
\begin{equation}
	\rho_b(r)=-\frac{1}{\pi} \int_r^\infty\frac{d\Sigma_b (x)}{dx} \frac{1}{\sqrt{x^2-r^2}}dx\,,
\end{equation}
so that, the rotational velocity of the galaxy bulge is obtained as
\begin{equation}
    v_b^2(r)=\dfrac{4\pi G}{r}\int_0^r \rho_b(r'){r'}^2 dr'\,.
    \label{eq:vel_bulge}
\end{equation}

The galactic disk can be approximated through an exponential profile for the surface mass density \cite{1959HDP....53..311D,Freeman:1970mx}:
\begin{equation}
	\Sigma_d(r)=\Sigma_{d,0}\, e^{-\frac{r}{a_d}}\,,
\end{equation}
with $\Sigma_{d,0}$ being its central value, and $a_d$ the characteristic scale radius. 
The rotational velocity of the disk component is then \cite{Freeman:1970mx}
\begin{equation}
    v_d^2(r)= \frac{\pi G \Sigma_{d,0}}{a_d}r^2 \left[I_0\left(X_d\right) K_0\left(X_d\right)-I_1\left(X_d\right) K_1\left(X_d\right)
			\right],
   \label{eq:vel_disk}
\end{equation}
where $X_d\equiv r/a_d$, while $I_i$ and $K_i$ are modified Bessel functions of the first and second kind, respectively.

As regards the structure of the dark halo, one may assume the NFW profile \cite{Navarro:1995iw}:
\begin{equation}
    \rho_h(r)=\frac{\rho_0}{X_h\left(1+X_h\right)^2}\,,
\end{equation}
with $\rho_0$ and $h$ being the dark halo scale density and radius, respectively.
Thus, the rotational velocity of the dark halo reads
\begin{equation}
    v_h^2(r)=  \frac{4\pi G}{r} \rho_0 h^3 \left[\ln(1+X_h)-\frac{X_h}{1+X_h}\right],
    \label{eq:vel_dark_halo}
\end{equation}
where we have defined the quantity $X_h\equiv r/h$.

\subsection{Rotation curves in Yukawa cosmology}

One can easily check that the modified Newtonian dynamics can explain the flat rotation curves of galaxies. Having the potential, we can obtain the circular speed of
an orbiting test object using 
\begin{eqnarray}
    v^2(r)=r \frac{d \Phi(r)}{dr}.
\end{eqnarray}
If we define the Newtonian potential \cite{Capozziello:2017rvz}
\begin{eqnarray}
    \Phi_N(r)=-\frac{G M(r)}{r},
\end{eqnarray}
and 
\begin{eqnarray}
    v_N^2(r)=r \frac{d \Phi_N(r)}{dr}=-\frac{G M(r)}{r}\,,
\end{eqnarray}
we obtain
\begin{equation}
v^2 = \frac{G M(r)}{r}+ \frac{ \alpha G M(r) }{r}\left(\frac{r+\lambda}{\lambda}\right)e^{-\frac{r}{\lambda}}.
\end{equation}
Near the galactic center, the first term dominates the total force, hence we can assume $M(r)=M_b(r)+M_d(r)$ as a sum of the bulge and disk masses, while the second term is important and can be attributed to the dark matter effect, which implies flat rotation curves of galaxies \cite{Jusufi:2023xoa}, namely
\begin{equation}
v^2 = \frac{ G M_b(r) }{r}+ \frac{ G M_d(r) }{r}+\frac{ G M_{DM}(r) }{r},
\end{equation}
where we have defined 
\begin{eqnarray}
   M_{DM}(r)= \alpha M(r) \left(\frac{r+\lambda}{\lambda}\right)e^{-\frac{r}{\lambda}}.
\end{eqnarray}
This gives the expression for the apparent dark matter that appears due to the modification of the gravitational potential. Thus, one can write the total velocity as
\begin{eqnarray}
v^2(r) &=& v_b^2(r)+v_d^2(r)+v^2_{DM}(r).
\end{eqnarray}
where $v_b^2(r)$ and $v_d^2(r)$ are given by Eqs.~\eqref{eq:vel_bulge} and \eqref{eq:vel_disk}, respectively, whereas the apparent dark matter contribution is
\begin{equation}
  v^2_{DM}=\frac{G M_{DM}(r)}{r} =\frac{ \alpha G M(r) }{r}\left(\frac{r+\lambda}{\lambda}\right)e^{-\frac{r}{\lambda}}\,,
\end{equation}
with $M(r)=M_b(r)+M_d(r)$ being the baryonic mass profile of the galaxy. The final expression for the total velocity can be then written as
\begin{equation}
    v=\sqrt{(v_b^2+v_d^2)\left[1+\dfrac{\alpha(r+\lambda)}{\lambda}e^{-\frac{r}{\lambda}}\right]}\,.
\end{equation}

We can further re-obtain the MOND expression found by Milgrom,  $v^4=GM a_0$ \cite{Milgrom:1983ca, Milgrom:1983pn, Milgrom:1983zz}. In our case, we have
\begin{equation}
    v^2 \simeq \frac{G M(r)}{r} + \sqrt{GM(r) \left[\frac{GM(r) (r+\lambda)^2 \alpha^2}{r^2\lambda^2}\right]e^{-\frac{2r}{\lambda}}}\,.
\end{equation}
In the outer part of the galaxy, the contribution of the first term is small, hence 
\begin{equation}
    v^2 \simeq  \sqrt{GM(r) \left[\frac{GM(r) (r+\lambda)^2 \alpha^2}{r^2\lambda^2}\right]e^{-\frac{2r}{\lambda}}},
\end{equation}
or 
\begin{equation}
    v^4 \simeq  GM(r) \left[\frac{GM(r) (r+\lambda)^2 \alpha^2}{r^2\lambda^2}\right]e^{-\frac{2r}{\lambda}}.
\end{equation}
This implies that the quantity
\begin{equation}
    a_0= \frac{GM\alpha^2}{\lambda^2}\left(1+\frac{\lambda}{r}\right)^2e^{-\frac{2r}{\lambda}}
\end{equation}
is responsible for the MOND acceleration which is observed to be $a_0 \sim 1.2 \times 10^{-10}\, \text{m/s}^2$. In fact, we can obtain this value in the outer part of the galaxy by taking the limit 
\begin{equation}
    \lim_{r \to \lambda}a_0= \frac{4GM\alpha^2}{\lambda^2}e^{-2}\simeq 1.2\,\times 10^{-10}\, \rm{m/s}^2\,,
\end{equation}
where $M \sim 1.2 \times 10^{40}\,\text{kg}$, and $\lambda \sim 0.74\,\text{kpc}$.  Therefore, contrary to the $\Lambda$CDM case, the rotation curve in our model does not need the dark matter halo contribution, but only the ones from the bulge and the disk. In the Yukawa scenario, dark matter may be thus seen as an apparent effect emerging from the modification of the gravitational potential. See also \cite{Bernal:2011qz,Capozziello:2017rvz}.

\subsection{The core-cusp problem and the role of NFW profile in Yukawa cosmology}

The core-cusp problem in $\Lambda$CDM cosmology is a well-known issue, particularly concerning small-scale cosmological phenomena within galactic centers \cite{Shi:2021tyg}. From an observational perspective, there is a preference for a constant dark matter density within the inner regions of galaxies. However, numerical simulations conducted within the framework of $\Lambda$CDM cosmology indicate a steep power-law-like behavior in the galactic center, suggesting the presence of dark matter spikes. These predicted dark matter spike regions, however, have never been observed, posing a significant challenge to the validity of the $\Lambda$CDM model.
To address this discrepancy, we can explore how a similar behavior may arise within the framework of Yukawa cosmology.

In the $\Lambda$CDM model, the NFW profile is very often used to describe the distribution of dark matter in the halos of galaxies and in cosmological simulations. However, since we have here an equation for the apparent dark matter obtained from the modified law of gravity, we should verify whether one can obtain effectively dark matter velocity due to the NFW from the Yukawa model. 

In particular, for the mass function in general we have $M(r)=M_{\rm star}(r)+M_{\rm gas}(r)+...$ including all the baryonic matter components, such as bulge, disk, dust, black hole, etc. The corrections to $\alpha$ start to appear at relatively large distances from the galactic center. Viewed from this distance, we can choose the following mass profile
\begin{eqnarray}
    M(r)=M\left(\frac{r}{r+r_0}\right),
\end{eqnarray}
where $M$ is some constant with dimensions of mass, and $r_0$ is a given scale radius. Specifically, we introduce the average apparent dark matter inside a region of volume $V$ through the relation
\begin{equation}
    \left \langle M_{DM}(r)\right \rangle= \frac{1}{V} \int_0^\pi \int_0^{2\pi}\int_0^{r}  M_{DM}(r') r'^2 d^3r'.
\end{equation}
We shall now investigate the region $r\ll \lambda$, that is, the region inside the galaxy. In this range, we have $e^{-r/\lambda}\simeq 1-r/\lambda$, then by solving the integral we obtain
\begin{align}
    \left \langle M_{DM}(r)\right \rangle & \simeq   4 \pi \alpha \rho_0  r_0^3 \left[\ln \left( 1+\frac{r}{r_0} \right) \left(\frac{r_0^2}{\lambda^2}-1\right)\right] \nonumber 
    \\ 
    &- 4 \pi \alpha \rho_0  r_0^3  \left[\frac{r (2 r_0-r)}{2\lambda^2} \right] +f(\alpha,\rho_0,r_0,r)\,,
\end{align}
where $\rho_0=M/V$, $r_0>\lambda$ and $r<\lambda$ and $f(\alpha,M,r_0,r)$ is some aribitrary function. For the average velocity in the region $r\ll \lambda$, due to the apparent dark matter, we get 
\begin{eqnarray}\notag
    \left \langle v^2_{DM}(r)\right \rangle & \simeq &  \frac{4 G \pi \alpha \rho_0  r_0^3}{r} \left[\ln \left( 1+\frac{r}{r_0} \right) \left(\frac{r_0^2}{\lambda^2}-1\right)\right] \\\notag
    &-& \frac{4 G \pi \alpha \rho_0  r_0^3}{r}  \left[\frac{r (2 r_0-r)}{2\lambda^2} \right]\\
    &+& \frac{4 G  f(\alpha,\rho_0,r_0,r)}{r}\,.
\end{eqnarray}
Notice that, as we pointed out earlier, we have only an apparent mass and not a real mass. In fact, the above expression looks similar to the dark matter velocity computed by using the NFW profile, suggesting that the latter could be an effective description of the Yukawa dark matter profile. 
In Yukawa cosmology, the core-cusp problem in the galactic center is alleviated. This issue arises when considering the average mass for apparent dark matter. However, it is crucial to recognize that this apparent mass is not an actual form of matter; rather, it emerges due to the distribution of baryonic matter and the modifications introduced by Yukawa's law of gravity.

\section{Observational constraints}
\label{sec:constraints}

\begin{table}
\begin{center}
\setlength{\tabcolsep}{1em}
\renewcommand{\arraystretch}{2}
\begin{tabular} {l|c c}
\hline
Parameter & MW & M31 \\
\hline
$\Sigma_{b,0}\,[\times 10^9 M_\odot]$ & $0.37\pm 0.03\,(0.06)$ & $0.14^{+0.02\,(0.08)}_{-0.05\,(0.06)}   $\\
$a_b$\,[kpc] & $2.20^{+0.10\,(0.21)}_{-0.11\,(0.19)}$ & $7.18^{+1.86\,(2.77)}_{-1.47\,(3.02)}$  \\
$\Sigma_{d,0}\,[\times 10^9 M_\odot]$ & $0.66\pm 0.02\,(0.05)$ & $0.43^{+0.08\,(0.21)}_{-0.12\,(0.19)}    $\\
$a_d$\,[kpc] & $9.05\pm 0.25\,(0.50)$ & $9.65^{+1.18\,(3.67)}_{-2.12\,(2.94)}$\\
$\alpha$ & $0.40^{+0.06\,(0.14)}_{-0.07\,(0.12)}$ & $0.37^{+0.11\,(0.31)}_{-0.17\,(0.25)}$\\
$\lambda$\,[kpc] & $0.74^{+0.05\,(0.13)}_{-0.07\,(0.11)}$ & $0.52^{+0.18\,(0.41)}_{-0.24\,(0.36)}      $\\
\hline
\end{tabular}
 \caption{68\% (95\%) C.L. estimates on the free parameters of the Yukawa scenario, as a result of the MCMC analysis of the MW and M31 galaxy rotation data.}
    \label{tab:MOND_results}
\end{center}
\end{table}

\begin{table}
\centering
\renewcommand{\arraystretch}{2}
\begin{tabular} { l| c c }
\hline
 Parameter &  MW & M31\\
\hline
$\Sigma_{b,0}\,[\times 10^9 M_\odot]$ & $0.66\pm 0.02\, (0.04) $ & $0.70^{+0.07\,(0.15)}_{-0.09\,(0.14)}$ \\
$a_b$\,[kpc] & $1.60\pm 0.05\,(0.09)$ & $1.31^{+0.14\,(0.31)}_{-0.19\,(0.28)}      $\\
$\Sigma_{d,0}\,[\times 10^9 M_\odot]$ & $0.61\pm 0.02\,(0.35)$ & $0.61\pm 0.17\,(0.32)      $\\
$a_d$\,[kpc] & $8.23 \pm 0.33\,(0.64)$ & $4.52^{+0.93\,(1.89)}_{-0.93\,(1.78)}           $\\
$\rho_0\,[M_\odot\,\text{kpc}^{-3}]$ & $0.71^{+0.15\,(0.34)}_{-0.19\,(0.31)}\times 10^4 $  & $1.70^{+0.53\,(1.45)}_{-0.98\,(1.19)}\times 10^7 $\\
$h\ [\text{kpc}]$ & $1.07\pm 0.15\,(0.34)\times 10^3 $  & $14.8^{+2.2\,(7.2)}_{-4.1\,(5.7)} $      \\
\hline
\end{tabular}
\caption{68\%\,(95\%) C.L. estimates on the free parameters of the $\Lambda$CDM model, as a result of the MCMC analysis of the MW and M31 galaxy rotation data.}
    \label{tab:LCDM_results}
\end{table}

\begin{figure*}
    \centering
    \includegraphics[width=3.3in]{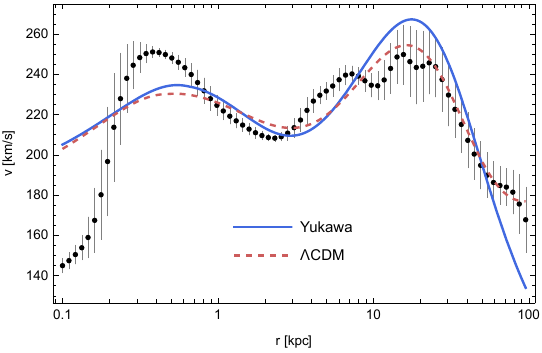}\qquad 
    \includegraphics[width=3.3in]{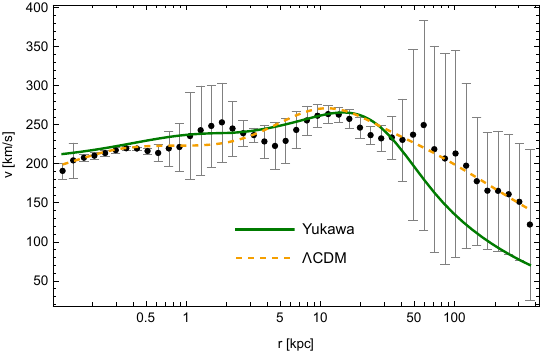}
    \caption{Galaxy rotation curves of the Yukawa scenario and the $\Lambda$CDM model, based on the mean results of the MCMC analysis of the MW (left panel) and M31 (right panel) measurements.}
    \label{fig:MW_comparison}
\end{figure*}

Let us now use galaxy rotation velocity data to constrain the theoretical scenario inferred from the Yukawa cosmology. Specifically, we consider the measurements at $r<100\, \text{kpc}$ obtained in \cite{Sofue:2020rnl} for the MW, and the measurements at $r<400\, \text{kpc}$ presented in \cite{Sofue:2015xpa} for the M31 galaxy. Then, we compare our findings with the predictions of the $\Lambda$CDM model built upon the NFW dark matter profile.

In the case of the Yukawa scenario, the set of parameters to be fitted is
\begin{equation}
    \bm{\theta}_\text{Yukawa}=\{\Sigma_{b,0},\, a_b,\, \Sigma_{d,0}, \,\alpha, \,\lambda\}\,,
\end{equation}
while, for the $\Lambda$CDM model, we have 
\begin{equation}
    \bm\theta_{\Lambda\text{CDM}}=\{\Sigma_{b,0},\, a_b,\, \Sigma_{d,0}, \,\rho_0, \,h\}\,.
\end{equation}
It is worth noticing that the number of free parameters is the same for both models.
To obtain observational bounds on the above sets, we consider the Likelihood function
\begin{equation}
    \mathcal{L}(\bm\theta)\propto e^{-\frac{1}{2}\chi^2(\bm\theta)}\,,
\end{equation}
where
\begin{equation}
    \chi^2(\bm\theta)=\sum_{i=1}^N \left[\dfrac{v_i-v(r_i,\bm\theta)}{\sigma_i}\right]^2\,.
\end{equation}
Here, $N$ is the number of rotation velocity measurements, $v_i$, in the galaxy catalogs, each with standard deviation $\sigma_i$, whereas $v(r_i,\bm\theta)$ is the theoretical prediction for the rotation velocity obtained in the context of the Yukawa scenario and $\Lambda$CDM model.

We thus applied a Markov Chain Monte Carlo (MCMC) analysis by means of the Metropolis-Hastings algorithm \cite{Hastings:1970aa} to sample the parameter space. 
Assuming flat priors on the fitting parameters, we use the Python package \texttt{getdist} \cite{Lewis:2019xzd} to analyze the chains. 

Our numerical results are shown in Tables~\ref{tab:MOND_results} and \ref{tab:LCDM_results} for the Yukawa scenario and the $\Lambda$CDM model, respectively. Specifically, for each parameter, we report the mean value, together with the $1\sigma$ and $2\sigma$ confidence level (C.L.), resulting from the MCMC analysis on the MW and M31 data.
In particular, we use the mean results to highlight in Figs.~\ref{fig:MW_comparison} the differences among the Yukawa rotation curves and those predicted by the standard $\Lambda$CDM paradigm.
Furthermore, Figs.~\ref{fig:MW_MOND} and \ref{fig:MW_LCDM} show the $1\sigma$ and $2\sigma$ C.L. regions and the posterior distribution for the Yukawa and $\Lambda$CDM parameters, respectively, obtained from the analysis of the MW measurements. The same quantities in the case of the M31 measurements are shown in Figs.~\ref{fig:M31_MOND} and \ref{fig:M31_LCDM}.

\subsection{Bayesian model selection}
\label{subsec:Bayesian}

Useful tools to measure the statistical performance of models are the Bayesian information criteria \cite{Kunz:2006mc,Liddle:2007fy}. In this regard, well-known examples are offered by the Akaike Information Criterion (AIC) \cite{AIC} and Bayesian Information Criterion (BIC) \cite{BIC}, which describe the effective model complexity by taking into account the number of free parameters that characterizes different theoretical scenarios\footnote{Applications of Bayesian information criteria in various cosmological contexts may be found, e.g., in Refs.~\cite{DAgostino:2018ngy,DAgostino:2019wko,Capozziello:2020ctn,DAgostino:2020dhv}.}.

However, since the models subject to the present study have the same number of free parameters, the AIC and BIC criteria reduce to the maximum likelihood estimation. According to the latter, for a given dataset, models characterized by the smaller values of $\chi^2$ are statistically favored with respect to those with a higher $\chi^2$ value. In our case, we can define the quantity
\begin{equation}
    \Delta\chi^2=\chi^2_\text{Yukawa}-\chi^2_{\Lambda\text{CDM}}\,,
\end{equation}
whose sign suggests which model is better performing, i.e., the Yukawa scenario for $\Delta\chi^2<0$, or the $\Lambda$CDM model in the case of $\Delta\chi^2>0$.

Additionally, one may consider the more powerful and accurate Deviance Information Criterion (DIC) \cite{Spiegelhalter:2002yvw,Trotta:2008qt}:
\begin{equation}
    \text{DIC}\equiv 2\langle-2\ln \mathcal{L}\rangle+2\ln\langle \mathcal{L} \rangle\,,
\end{equation}
where $\langle \cdot \rangle$ denotes a mean over the posterior distribution. 
The main advantage of the DIC with respect to the AIC and BIC is that it involves an effective number of degrees of freedom, thus accounting for parameters that are unconstrained by the data. As seen for the $\chi^2$ case, to measure the goodness of the model fittings, we analyze the difference
\begin{equation}
    \Delta\text{DIC}=\text{DIC}_\text{Yukawa}-\text{DIC}_{\Lambda\text{CDM}}\,.
\end{equation}

We summarize our results in Table~\ref{tab:selection}. The highly negative values of both statistical indicators suggest a strong preference for the Yukawa scenario over the $\Lambda$CDM model when used to fit the MW data. Conversely, in the case of the M31 data, both $\Delta\chi^2$ and $\Delta$DIC are positive, indicating that the standard $\Lambda$CDM model is significantly favored with respect to the Yukawa scenario inferred from the Yukawa cosmology. 
As we argue in the Appendix, this issue is actually related to the precision of the data employed in the analysis. In fact, we demonstrate that utilizing a more recent and accurate sample of measurements can significantly improve the predictive power of the Yukawa scenario in the outer galaxy regions, thereby reducing the gap in the Bayes factor compared to the $\Lambda$CDM model.

\begin{table}
    \centering
    \setlength{\tabcolsep}{1em}
\renewcommand{\arraystretch}{2}
    \begin{tabular}{c|c c}
    \hline
    Data & $\Delta\chi^2$ & $\Delta$DIC\\
    \hline
       MW  & $-28.9$  & $-27.5$\\
       M31 & $11.0$ & $10.7$ \\
    \hline
    \end{tabular}
    \caption{Bayesian model selectors as a result of the MCMC analysis of the MW and M31 data. The differences $\Delta$ are calculated with respect to the $\Lambda$CDM model.}
    \label{tab:selection}
\end{table}

\section{Discussion}
\label{sec:discussion}

Let us discuss now our previous results in light of some recent findings in the literature. To do so, we restore the SI units of measurement.

In particular, in Refs.~\cite{Gonzalez:2023rsd,Jusufi:2023xoa} it has been shown that the Compton wavelength of the graviton suggested by cosmological data is of the order of Gpc, specifically 
\begin{eqnarray}
    \lambda^{\rm cosmology} \sim 10^{26}\, \rm{m}\,.
\end{eqnarray}
In the present paper, however, we find that $\lambda$ is of the order of kpc, specifically,
\begin{eqnarray}
    \lambda^{\rm galaxy} \sim 10^{19}\, \rm{m}\,,
\end{eqnarray}
in agreement with graviton bounds coming from gravitational waves observations \cite{LIGOScientific:2016aoc}.
It is then natural to ask how one can reconcile this apparent inconsistency. As we shall elucidate below, the response to this question is surprisingly concise and remarkable. Specifically, the graviton possesses an exceedingly small mass, namely
\begin{eqnarray}
    m_g=\frac{\hbar}{\lambda c}\,,
\end{eqnarray}
so that, it is imperative for it to have a quantum mechanical description, a factor that could prove significant even on cosmological scales.  Hence, an uncertainty principle for the graviton must exist:
\begin{eqnarray}
    \Delta p\, \Delta x \sim \hbar\,.
\end{eqnarray}
Usually, the uncertainty in measurements arising from fundamental limitations due to the Heisenberg principle is a quantum concept, but it possesses interesting implications also on cosmological scales \cite{Capozziello:2020nyq}. At large scales, in fact, we have more uncertainty in position $\lambda^{\rm cosmology} =\Delta x^{\rm cosmology} \sim 10^{26}$ m,  but more precision on the momentum $\Delta p^{\rm cosmology}$, namely 
\begin{eqnarray}
     \Delta p^{\rm cosmology} \sim \frac{\hbar}{\Delta x^{\rm cosmology}} \sim 10^{-60}\,\rm{N}\cdot\rm{s}\,.
\end{eqnarray}
However, since $\Delta p^{\rm cosmology}=m_g c$, the graviton mass results to be $m_g=10^{-68}\,\text{kg}$, and consequently we get 
\begin{eqnarray}
     \lambda^{\rm cosmology}=\frac{\hbar}{m_g c}\sim 10^{26}\, \rm{m}\,.
\end{eqnarray}
In other words, applied to the whole universe, we have more uncertainty in position but more precision on the momentum and hence on the graviton mass. 

On the other hand, applied to the galaxy scales, we have less uncertainty in position $\lambda^{\rm galaxy} =\Delta x^{\rm galaxy} \sim 10^{19}$ m,  but more uncertainty in momentum $\Delta p^{\rm galaxy}$, namely
\begin{eqnarray}
     \Delta p^{\rm galaxy} \sim \frac{\hbar}{\Delta x^{\rm galaxy}} \sim 10^{-53}\,\rm{N}\cdot\rm{s}\,.
\end{eqnarray}
By making use of $\Delta p^{\rm galaxy}=m_g c$, we find the graviton mass $m_g=10^{-61} $kg, and consequently 
\begin{eqnarray}
     \lambda^{\rm galaxy}=\frac{\hbar}{m_g c}\sim 10^{19} \rm{m}\,.
\end{eqnarray}
This explains perfectly well the discrepancies between our results and the findings of \cite{Gonzalez:2023rsd,Jusufi:2023xoa}. 
In fact, as pointed out in Ref.~\cite{Jusufi:2024ifp}, the graviton wavelength $\lambda$ is, in general, a function of the redshift, i.e., $\lambda(z)$, and the graviton mass can fluctuate with the cosmological scales. It follows that
\begin{eqnarray}
    \lambda(z) \sim \frac{\hbar}{m_g(z)\, c}.
\end{eqnarray}
By differentiating the last equations and by using $d \lambda \simeq \Delta \lambda$ along with $d m_g \simeq \Delta m_g$, we can  write
\begin{eqnarray}
  \Bigg|\frac{\Delta \lambda}{\lambda}\Bigg| \sim \Bigg|\frac{\Delta m_g}{m_g}\Bigg|\,.
\end{eqnarray}
Applying this relation to the cosmological and galactic scales, we have 
\begin{equation}
  \frac{\lambda^{\rm cosmology}-\lambda^{\rm galaxy}}{\lambda^{\rm galaxy}}\sim\frac{m_g^{\rm galaxy}-m_g^{\rm cosmology}}{m_g^{\rm cosmology}}\sim 10^7,
\end{equation}
which resolves the apparent inconsistency between the two measurements.
To summarize, both constraints are correct, however, there are fundamental limitations of measurements even in cosmology, and that is the reason for the difference found in the analyses of galactic and cosmological scales. 
The mismatch between the two scales may arise due to screening mechanisms, such as the chameleon mechanism \cite{Khoury:2003rn}. Specifically, the latter could result in a graviton mass that varies according to the surrounding environment. As for the case of theories of massive bigravity \cite{DeFelice:2017oym,DeFelice:2017gzc}, the increased matter density within galactic scales leads to a particle mass with an interaction range of the order of kpc, whereas, 
in cosmological contexts with significantly lower matter density, the graviton mass is smaller, resulting in an exceptionally long-range interaction, potentially on the scale of Mpc or Gpc. Similar ideas have been investigated in various contexts, e.g., Ref.~\cite{Aoki:2017ffl}.

\subsection{Implications to gravitational waves}

Finally, let us comment here about the possible implications of our results in terms of massive gravitons. This may prove useful in light of the recent results of the NANOGrav collaboration, which has shown the potential contribution of massive gravitons to gravitational waves in the $n$Hz frequency range \cite{NANOGrav:2023gor}. 
Specifically, the frequency associated with a massive graviton reads
\begin{eqnarray}
    \frac{\omega}{c}=\sqrt{\frac{m_g^2 c^2}{\hbar^2}+|\vec{k}|^2}\,,
\end{eqnarray}
with the four-wave vector $k^{\mu}\equiv(\omega/c,\vec{k})$. The above equation suggests that there exists a minimal frequency for the gravitational-wave signal due to the massive graviton, namely \cite{Wu:2023rib}
\begin{eqnarray}
    f_{\rm min}=\frac{m_g c^2}{2 \pi \hbar}.
\end{eqnarray}

Adopting the results from the galaxy rotation curves analyzed through the Yukawa scenario in the present paper, we can infer estimates of the graviton mass. In particular, from the MW data analysis, we find
\begin{align}
&m_g\simeq (1.54 \pm 0.12)\times 10^{-62}\, \text{kg}\quad (1\sigma)\\
&m_g\simeq (1.54 \pm 0.25)\times 10^{-62}\, \text{kg}\quad (2\sigma)
\end{align}
while, from the M31 data analysis, we get
\begin{align}
    &m_g\simeq (2.19 \pm 0.89)\times 10^{-62}\, \text{kg}\quad (1\sigma)\\
    &m_g\simeq (2.19 \pm 1.62)\times 10^{-62}\, \text{kg}\quad (2\sigma)
\end{align}
Using the best-fit values above, we obtain the following minimal frequencies:
\begin{align}
    &f_{\rm min} \simeq 2.09 \times 10^{-12}\, \rm{Hz}\,, \\
    & f_{\rm min} \simeq 2.97 \times 10^{-12}\, \rm{Hz}\,,
\end{align}
for the MW and M31, respectively. 

It is important to stress that the minimal value of the frequency depends on the graviton mass, however, we saw that there is a fundamental limitation in measuring the position (length scale) and the momentum (graviton mass). This means that different measurements point out different values of the momentum due to the different lengths of characteristic observation. This follows from the uncertainty principle, as previously argued. In particular, considering the length scale of large-scale structures to be $\Delta x \sim 10^{25}\, \text{m}$ implies that $m_g \sim 10^{-68}\,\text{kg}$, which gives rise to the minimal frequency $f_{\rm min}\sim 10^{-18}\, \text{Hz}$. For galactic scales of $\Delta x \sim 10^{19}\, \text{m}$, we saw that $f_{\rm min}\sim 10^{-12}$ Hz. Finally, if we consider the scale for the pulsar-timing array observations of the gravitational wave background with a characteristic length of $\Delta x \sim 10^{16}\, \text{m}$ \cite{Konoplya:2023fmh}, we have $m_g \sim 10^{-59}\, \text{kg}$, leading to $f_{\rm min}\sim 10^{-9}\, \text{Hz}$. Therefore, we can say that the range of frequency depends on the length of characteristic observation due to the uncertainty principle.  Recent observations at $n$Hz frequency do not forbid massive gravitons and it can be explained in terms of the scale of measurements due to the uncertainty principle. Usually, the observation of these frequencies, in the $n$Hz range, is linked to the gravity weaves produced by binary mergers of supermassive black holes \cite{Rajagopal:1994zj,Wyithe:2002ep,Sesana:2004sp}. However, other possibilities for a stochastic gravitational wave background have been investigated, including gravity waves from inflation (see Refs.~\cite{Vagnozzi:2023lwo,Correa:2023whf} and references therein), and the role of the graviton mass in gravitational waves with very long wavelengths \cite{Lee:2010cg,Wu:2023rib,Konoplya:2023fmh}.

\section{Conclusions and Outlook}
\label{sec:conclusion}

In this paper, we analyzed the galaxy rotation curves within the framework of the Yukawa corrected gravitational potential. Our research draws upon observational data regarding the rotational velocities observed in the MW and M31 galaxies. While employing the $\Lambda$CDM model, we thoroughly considered the contributions originating from the galactic bulge, galactic disk, and the NFW profile for the dark matter halo. On the other hand, using the modified gravitational force in the framework of the Yukawa model, we found analytical expressions characterizing the contribution of dark matter velocity.

To analyze the galaxy rotation data, we performed a Bayesian analysis based on the MCMC numerical technique. Specifically, we constrained the free parameter of the Yukawa model up to $2\sigma$ C.L., and we obtained the posterior distribution for each coefficient after marginalizing over the parameter space. Moreover, we compared our results to the predictions of the standard $\Lambda$CDM paradigm. In particular, we measured the statistical performance of our theoretical scenario through information criteria taking into account the effective number of degrees of freedom with respect to the $\Lambda$CDM reference.

Our findings notably show that the galactic rotation curves can be comprehensively explained without resorting to the presence of dark matter. Within the scope of Yukawa gravity, our study aligns with the recovery of MOND, suggesting that the existence of dark matter might be thought of as an apparent effect resulting from the modification of the law of gravitation. This modification is explicated in terms of the two key parameters of the Yukawa potential, namely the coupling constant $\alpha$, and the wavelength $\lambda$.

Furthermore, our research underscores a fundamental constraint in the precise determination of $\lambda$ due to the implications of quantum mechanics, particularly associated with the relatively small mass of the graviton. This constraint manifests differently across various scales. On cosmological scales, the uncertainty principle entails high uncertainty in position, but less uncertainty in $\lambda$ and the graviton mass. In contrast, on galactic scales, there exists a higher level of uncertainty in $\lambda$ and, consequently, in the graviton mass, with comparatively less uncertainty in position. This discrepancy elucidates the different orders of magnitude observed in cosmological data $(\lambda\sim \text{Gpc})$ and in galactic observations $(\lambda\sim \text{kpc})$. 

Finally, we used our MCMC results to constrain the graviton mass. Specifically,  we obtained the minimal frequency $f_{\rm min} \sim 10^{-12}\, \text{Hz}$  from both MW and M31 galaxy analyses. However, as we discussed, the frequency depends on the graviton mass, and in particular we saw that there is a fundamental limitation of measuring the position and the momentum of the graviton due to the uncertainty principle. This implies that, at scales corresponding to the pulsar-timing array observations, we can find the $n$Hz frequency, in accordance with the uncertainty principle.  Therefore, we expect massive gravitons to play an important role, in light of the recent observations of $n$Hz frequency obtained from the NANOGrav collaboration.

\acknowledgments

R.D. and S.C. acknowledge the financial support of the Istituto Nazionale di Fisica Nucleare (INFN) - Sezione di Napoli, \textit{inizative specifiche} QGSKY and MOONLIGHT2. This paper is based upon work from COST Action CA21136 - Addressing observational tensions in cosmology with systematics and fundamental physics (CosmoVerse), supported by COST (European Cooperation in Science and Technology).

\begin{widetext}

\begin{figure*}
    \centering
    \includegraphics[width=0.9\textwidth]{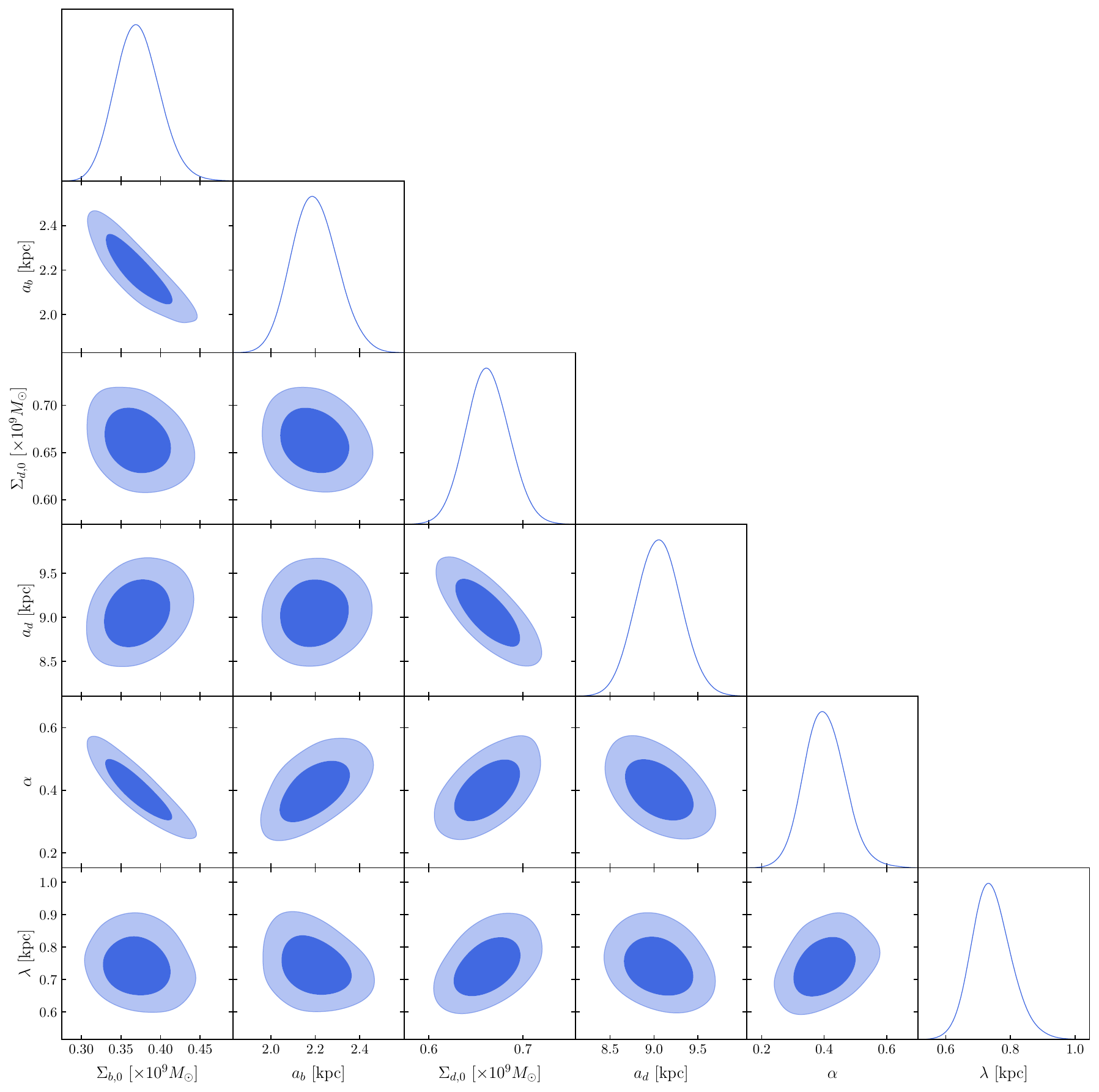}
    \caption{Marginalized 68\% and 95\% C.L. contours and posterior distributions for the free parameters of the Yukawa scenario, as a result of the MCMC analysis of the MW galaxy rotation data.}
    \label{fig:MW_MOND}
\end{figure*}

\begin{figure*}
    \centering
    \includegraphics[width=0.9\textwidth]{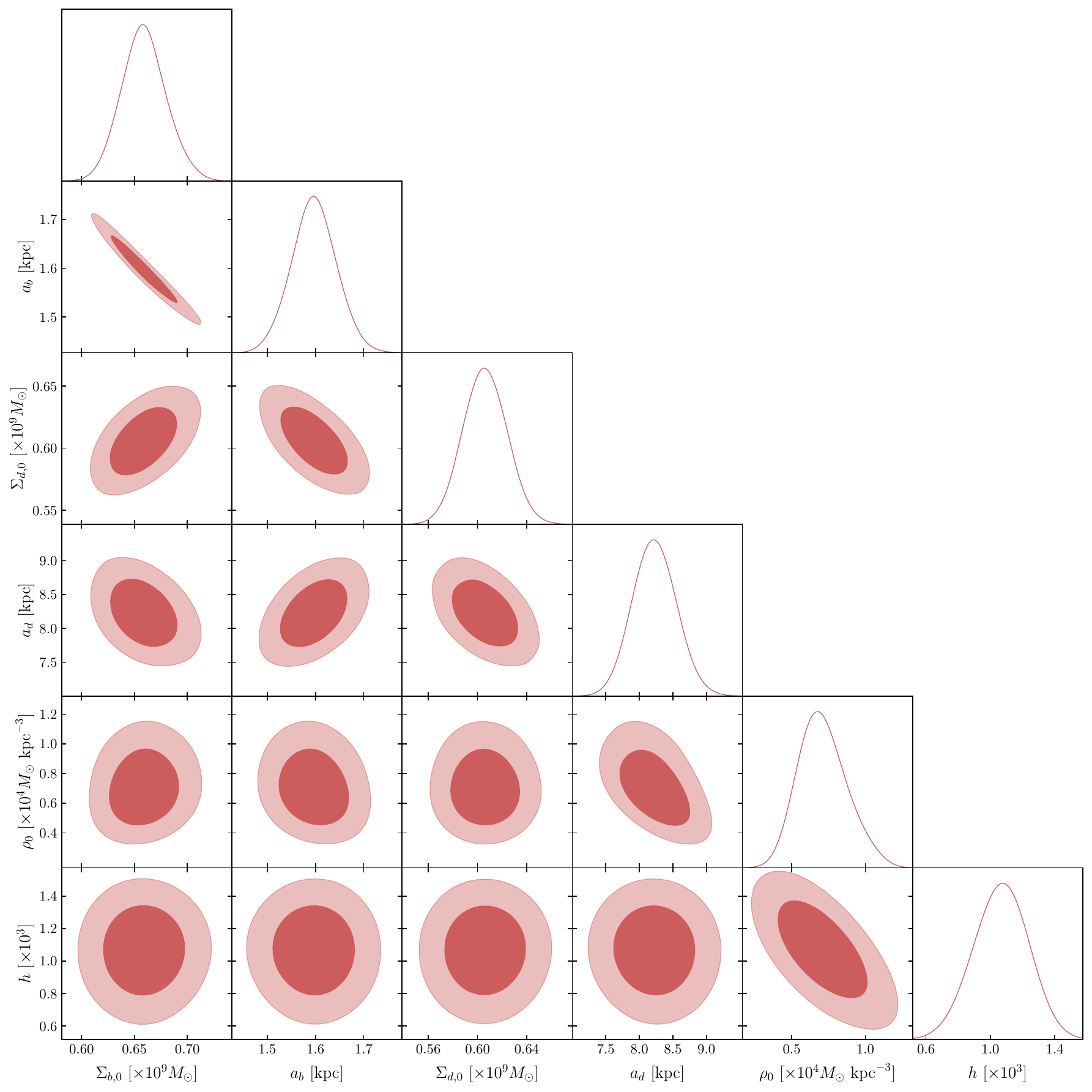}
    \caption{Marginalized 68\% and 95\% C.L. contours and posterior distributions for the free parameters of the $\Lambda$CDM model, as a result of the MCMC analysis of the MW galaxy rotation data.}
    \label{fig:MW_LCDM}
\end{figure*}

\begin{figure*}
    \centering
    \includegraphics[width=0.9\textwidth]{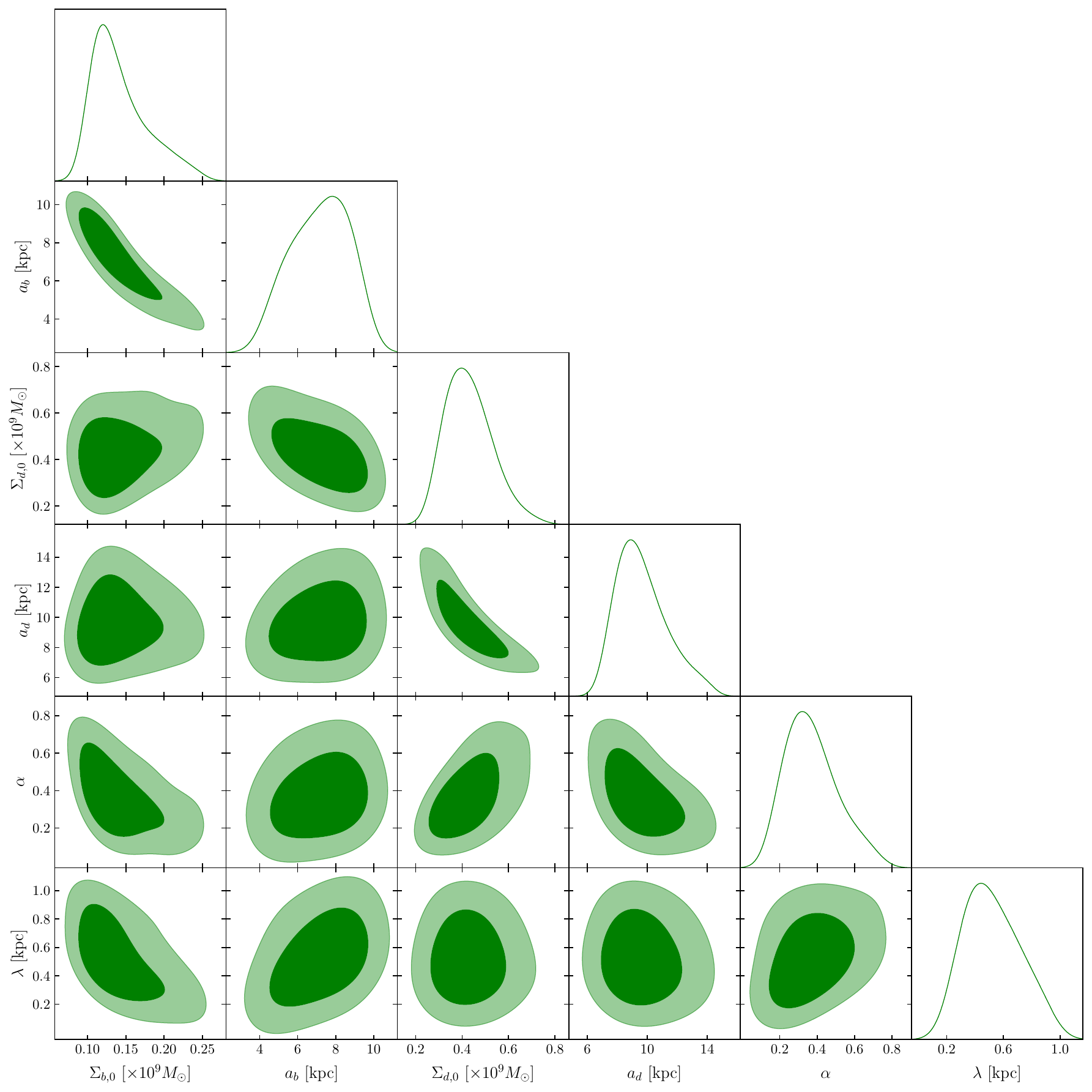}
    \caption{Marginalized 68\% and 95\% C.L. contours and posterior distributions for the free parameters of the Yukawa scenario, as a result of the MCMC analysis of the M31 galaxy rotation data.}
    \label{fig:M31_MOND}
\end{figure*}

\begin{figure*}
    \centering
    \includegraphics[width=0.9\textwidth]{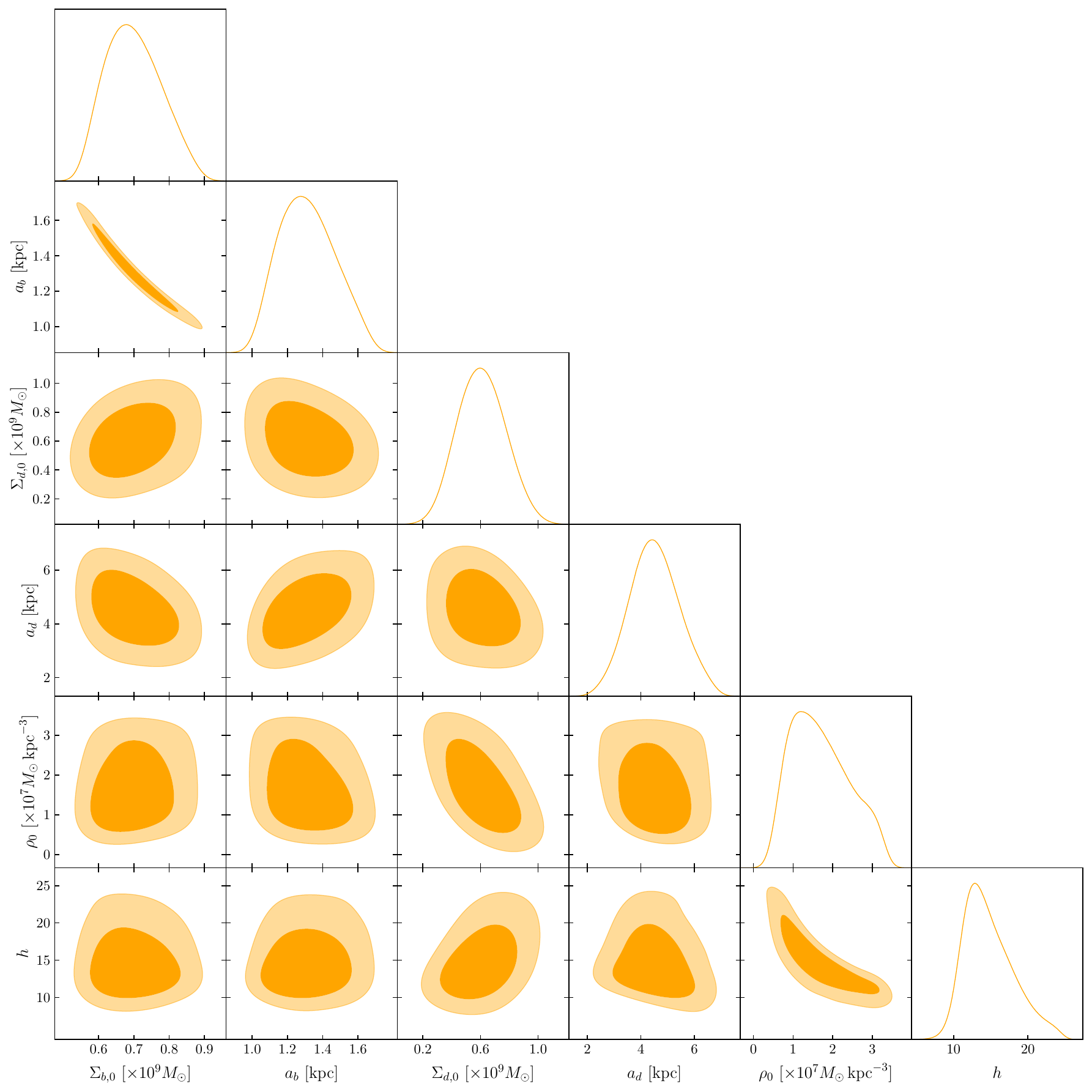}
    \caption{Marginalized 68\% and 95\% C.L. contours and posterior distributions for the free parameters of the $\Lambda$CDM model, as a result of the MCMC analysis of the M31 galaxy rotation data.}
    \label{fig:M31_LCDM}
\end{figure*}

\clearpage

\end{widetext}

\begin{widetext}

\appendix*
\section{}
\label{sec:appendix}

We here examine how the accuracy of galaxy data impacts the outcome of the numerical analysis. As a test, we can use the more accurate measurements of M31 recently provided in Ref.~\cite{2024MNRAS.528.2653Z}. Therefore, we performed an MCMC analysis on the new M31 dataset to constrain at the 68\% and 95\% C.L. the free parameters of the Yukawa scenario and the $\Lambda$CDM model. The new results are summarized in Table~\ref{tab:M31_results_new}. Based on the mean results of the MCMC analysis, we display the comparison between the predictions of both scenarios in Fig.~\ref{fig:M31_comparison_new}.
Moreover, in Figs.~\ref{fig:M31_MOND_new} and \ref{fig:M31_LCDM_new}, we show the marginalized $1\sigma$ and $2\sigma$ contours, along with the posterior distributions, for the model parameters.

To quantify the statistical performance of the theoretical scenarios, we employ Bayesian model selectors as described in Sec.~\ref{subsec:Bayesian}. In this case, we find
\begin{align}
    \Delta\chi^2=2.59\,, \quad \Delta\text{DIC}= 3.06\,,
\end{align}
indicating that the preference of the $\Lambda$CDM model over the Yukawa scenario is not decisive.
Furthermore, if we compare these results with those of the second row of Table~\ref{tab:selection}, we notice a reduction in the evidence against the Yukawa scenario. This provides further proof of the effectiveness of the Yukawa model in taking into account observations at all galactic distances.

\vspace{1cm}

\begin{table*}[h!]
\begin{center}
\setlength{\tabcolsep}{1em}
\renewcommand{\arraystretch}{2}
\begin{tabular} {l|c||l|c}
\hline
Parameter & Yukawa & Parameter & $\Lambda$CDM \\
\hline
$\Sigma_{b,0}\,[\times 10^8 M_\odot]$ & $0.70^{+0.13\,(0.41)}_{-0.23\,(0.35)} $ & $\Sigma_{b,0}\,[\times 10^8 M_\odot]$ & $1.77^{+0.36\,(0.96)}_{-0.58\,(0.80)}      $  \\
$a_b$\,[kpc] & $14.4^{+3.1\,(7.9)}_{-4.4\,(6.9)} $ & $a_b$\,[kpc] &  $4.8^{+0.8\,(2.7)}_{-1.6\,(2.0)}       $  \\
$\Sigma_{d,0}\,[\times 10^8 M_\odot]$ & $0.87^{+0.27\,(0.81)}_{-0.45\,(0.64)}      $ & $\Sigma_{d,0}\,[\times 10^8 M_\odot]$ & $0.32^{+0.12\,(0.40)}_{-0.27\,(0.30)}      $\\
$a_d$\,[kpc] & $28^{+7\,(20)}_{-10\,(20)}             $ & $a_d$\,[kpc] & $45^{+20\,(50)}_{-30\,(40)}$  \\
$\alpha$ & $0.13^{+0.05\,(0.13)}_{-0.10\,(0.12)}    $ & $\rho_0\,[\times 10^6 M_\odot\,\text{kpc}^{-3}]$ & $22^{+9\,(10)}_{-7\,(10)}              $ \\
$\lambda$\,[kpc] & $1.54^{+0.76\,(1.34)}_{-0.88\,(1.31)}$ & $h\ [\text{kpc}]$ &  $10.7^{+1.4\,(4.6)}_{-2.4\,(3.3)}        $\\
\hline
\end{tabular}
 \caption{MCMC results for the Yukawa  and the $\Lambda$CDM models using the new M31 dataset.}
    \label{tab:M31_results_new}
\end{center}
\end{table*}

\begin{figure*}
   \begin{center}
    \includegraphics[width=3.5in]{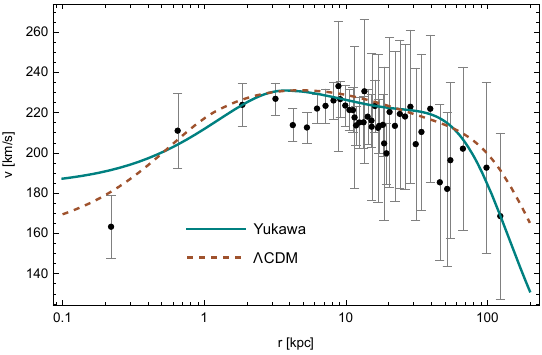}
    \caption{Rotation curves of the Yukawa scenario and the $\Lambda$CDM model, based on the mean results of the MCMC analysis of the new M31 dataset.}
    \label{fig:M31_comparison_new}
    \end{center}
\end{figure*}

\begin{figure*}
    \centering
    \includegraphics[width=0.9\textwidth]{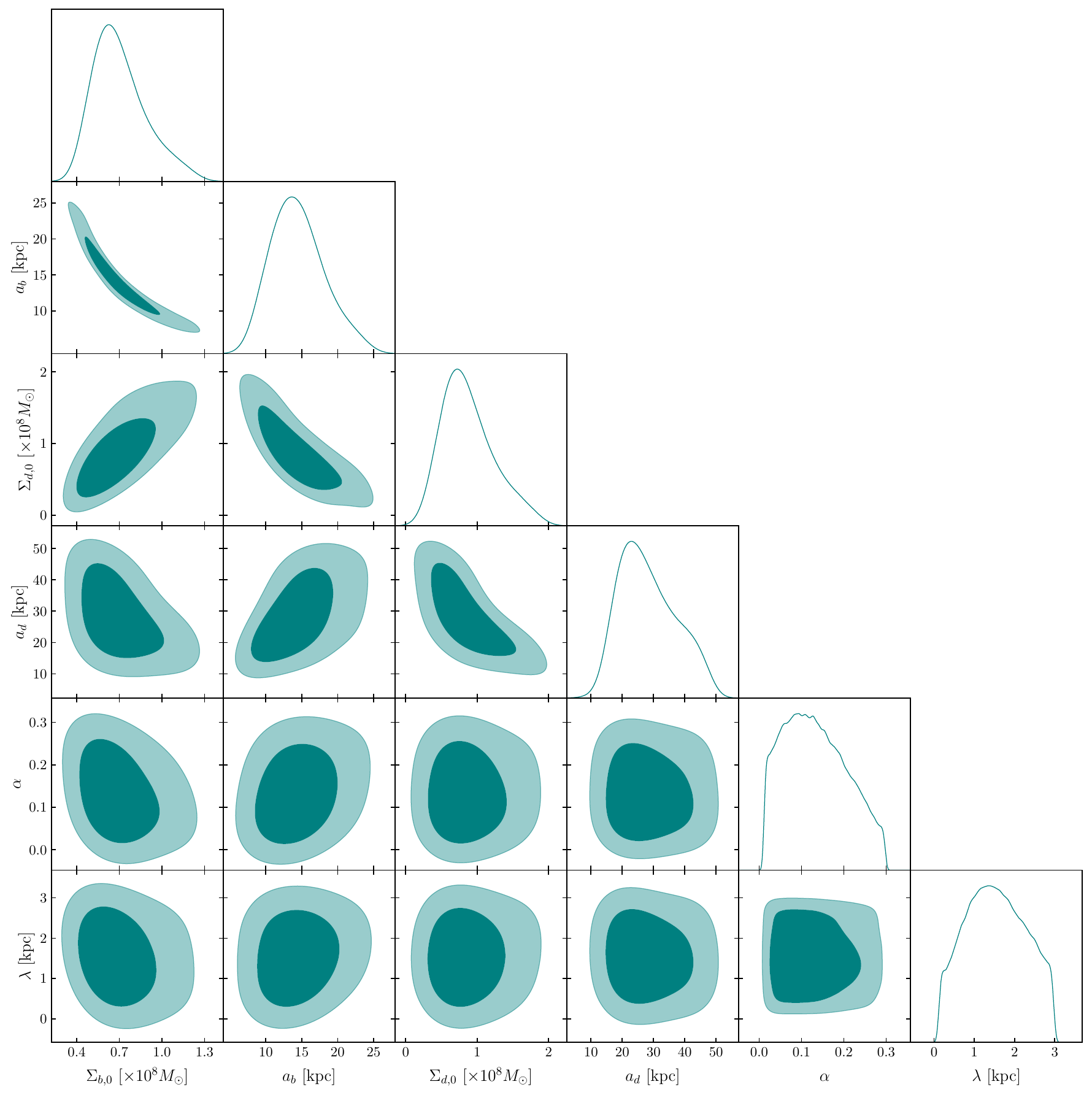}
    \caption{Contours for the Yukawa model using the new M31 data.}
    \label{fig:M31_MOND_new}
\end{figure*}

\begin{figure*}
    \centering
    \includegraphics[width=0.9\textwidth]{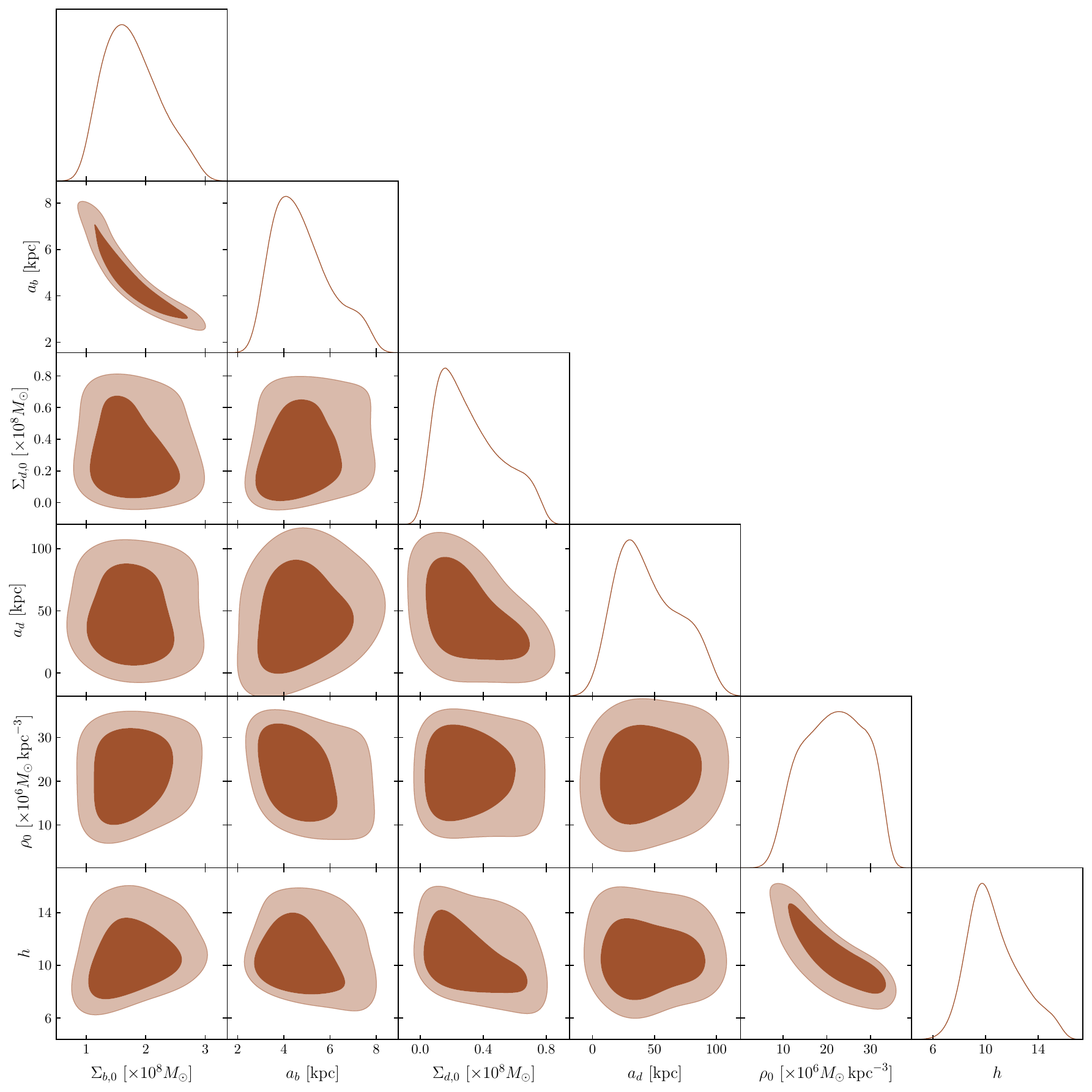}
    \caption{Contours for the $\Lambda$CDM model using the new M31 data.}
    \label{fig:M31_LCDM_new}
\end{figure*}

\clearpage

\end{widetext}

\bibliography{references}

\end{document}